\documentstyle[11pt,aaspp4]{article}
\def\la{\mathrel{\hbox{\rlap{\hbox{\lower4pt\hbox{$\sim$}}}\hbox{$<$}}}}
\def\ga{\mathrel{\hbox{\rlap{\hbox{\lower4pt\hbox{$\sim$}}}\hbox{$>$}}}}
\def\lesssim{\mathrel{\hbox{\rlap{\hbox{\lower4pt\hbox{$\sim$}}}\hbox{$<$}}}}
\def\etal{et al.\,\,}

\slugcomment{Accepted to {\it The Astrophysical Journal}}

\begin{document}

\title{Optical and Near Infrared Observations of the Afterglow of GRB 980329 from 15 Hours to 10 Days\footnote{Part of the data presented herein were obtained at the W. M. Keck Observatory, which is operated as a scientific partnership among the California Institute of Technology, the University of California and the National Aeronautics and Space Administration.  The Observatory was made possible by the generous financial support of the W. M. Keck Foundation.}}

\author{Daniel E. Reichart\altaffilmark{2}, Donald Q. Lamb\altaffilmark{2}, Mark R. Metzger\altaffilmark{3}, Jean M. Quashnock\altaffilmark{2}, \nl
David M. Cole\altaffilmark{2}, Francisco J. Castander\altaffilmark{2}, Sylvio Klose\altaffilmark{4}, James E. Rhoads\altaffilmark{5}, \nl
Andrew S. Fruchter\altaffilmark{6}, Asantha R. Cooray\altaffilmark{2}, and Daniel E. Vanden Berk\altaffilmark{7}}

\altaffiltext{2}{Department of Astronomy and Astrophysics, University
of Chicago, 5640 South Ellis Avenue, Chicago, IL 60637}

\altaffiltext{3}{Division of Physics, Mathematics, and Astronomy
105-24, Caltech, Pasadena, CA 91125}

\altaffiltext{4}{Th\"uringer Landessternwarte Tautenburg, Karl-Schwarzschild-Observatorium, D-07778 Tautenburg, Germany}

\altaffiltext{5}{Kitt Peak National Observatory, National Optical Astronomy Observatories, 950 North Cherry Avenue, Tucson, AZ 85719, which is operated by the Association of Universities for Research in Astronomy, Inc. (AURA) under cooperative agreement with the National Science Foundation}

\altaffiltext{6}{Space Telescope Science Institute, 3700 San Martin Drive, Baltimore, MD 21218}

\altaffiltext{7}{McDonald Observatory, University of Texas, RLM 15.308,
Austin, TX 78712}

\begin{abstract}
We report $I$-band observations of the GRB 980329 field made on March 29 with the 1.34-m Tautenberg Schmidt telescope, $R$-, $J$- and $K$-band observations made on April 1 with the APO 3.5-m telescope, $R$- and $I$-band observations made on April 3 with the Mayall 4-m telescope at KPNO, and $J$- and $K$-band observations made between April 6 - 8 with the Keck-I 10-m telescope. 
We show that these and other reported measurements are consistent with a power-law fading of the optical/near infrared source that is coincident with the variable radio source VLA J0702+3850.  This firmly establishes that this source is the afterglow of GRB 980329.  
\end{abstract}


\section{Introduction}

On March 29.16, GRB 980329 triggered the {\it BeppoSAX} Gamma Ray Burst Monitor and was detected simultaneously with the {\it BeppoSAX} Wide Field Cameras, which yielded an error circle of 3 arcmin radius (Frontera \etal 1998).  Seven hours later, this error circle was observed with the {\it BeppoSAX} Narrow Field Instruments, which detected a fading medium-energy X-ray source, 1SAX J0702.6+3850, about 1 arcmin from the center of the error circle (in 't Zand \etal 1998a).
    
On April 3, Taylor, Frail, \& Kulkarni (1998) reported the detection of a variable radio source, VLA J0702+3850, within the 1 arcmin radius error circle of 1SAX J0702.6+3850, with measurements made between March 30 and April 2.   On the same day, Djorgovski \etal (1998) reported the detection of a faint ($R = 25.7 \pm 0.3$ mag) optical source coincident with VLA J0702+3850, with measurements made on April 2 with the Keck-II 10-m telescope; they interpreted this source as the host galaxy of GRB 980329.   Also on the same day, Klose (1998) (see also Klose, Meusinger, \& Lehmann 1998) reported an $I$-band measurement of this source, made on March 30 with the Tautenburg Schmidt telescope, and Larkin \etal (1998a,b) reported $K$-band measurements of this source, made on April 2 and April 3 with the Keck-I 10-m telescope.

On April 4, Mannucci \etal (1998) reported a $J$-band measurement of a source, made on March 30 with the Gornergrat Infrared Telescope (TIRGO); however, the coincidence of their source with VLA J0702+3850 cannot be established (Palazzi \etal 1998b).  On April 5, Cole \etal (1998a,b) reported a $J$-band upper limit at the position of VLA J0702+3850, made on April 1 with the APO 3.5-m telescope, that, when compared to the $J$-band measurement of Mannucci \etal (1998), suggested that this source was fading.

Observations reported on April 6 by Palazzi \etal (1998a) ($R$-band, March 30), and on April 8 by Pedersen \etal (1998) ($R$-band, March 31 $+$ April 1 $+$ April 2) provided evidence that the source coincident with VLA J0702+3850 was fading.   On April 7, Smith \& Tilanus (1998a,b) reported the detection of a fading submillimeter source coincident with this source, with measurements made between April 5 and April 7 with SCUBA at the James Clerk Maxwell Telescope.  On April 12, Metzger (1998a,b) reported $K$-band measurements of the source, made between April 6 and April 8 with the Keck-I 10-m telescope, which demonstrated that the source continued to fade in the near infrared (NIR) as late as 10 days after the gamma-ray burst (GRB).  

Recently, Fruchter (1998b) reported that the source, as well as an underlying or host galaxy, is visible in a deep {\it HST}/NICMOS image, made approximately 200 days after the GRB.  The faintness of the underlying or host galaxy contradicts the previous claim of Djorgovski \etal (1998), unless the galaxy is extremely blue and/or is at very high redshift.

Except for the X-ray measurements, which can be found in in 't Zand \etal (1998b), the submillimeter measurements, which can be found in Smith \etal (1998), the radio measurements, which can be found in Taylor \etal (1998), and the NIR measurement of Fruchter (1998b), which is not yet available, all of the above measurements and upper limits, as well as the measurements and upper limits that we report in this paper, are listed in Table 1.  

In this paper, we report Tautenburg, Apache Point Observatory (APO), Kitt Peak National Observatory (KPNO), and Keck-I observations of the GRB 980329 field. 
In \S2, we report KPNO $R$- and $I$-band calibrations of the GRB 980329 field.  We report the Tautenburg observations in \S3, the APO observations in \S4, the KPNO observations in \S5, and the Keck-I observations in \S6.  In \S7, we discuss some of the implications of these and the other reported observations.  We summarize our results in \S8.

\section{KPNO $R$- and $I$-band Calibrations of the GRB 980329 Field}

On April 3, Rhoads, on behalf of the KPNO GRB followup team, observed the GRB 980329 field and the SA107 field of Landolt (1992) standard stars with the Mayall 4-m telescope at KPNO using the Mosaic CCD Imager.  Four 10-minute exposures were taken of the GRB 980329 field, two through the $R$-band filter and two through the $I$-band filter.  Four brief exposures were taken of the SA107 field, likewise, two through the $R$-band filter and two through the $I$-band filter. 
Weather was photometric, though image quality was poor, ranging from 1.25 to 1.75 arcsec, approximately.  The GRB 980329 field was observed at an airmass of 1.1, and the SA107 field was observed at an airmass of 1.4.  
All data reduction was done in IRAF, using the packages MSCRED for basic data reduction, and APPHOT and PHOTCAL for the photometric measurements and transformation equations.

Reliable fluxes were measured for the Landolt standard stars SA107-212, 213, 357, 359, 351, 457, 456, 600, 599, 612, 626, and 627.  Of these, the first five were measured twice in each filter, and the remainder were measured only once in each filter.  
Fluxes were corrected to a 7 arcsec radius aperture (the usual aperture size used by Landolt) using a curve of growth derived from all of the cleanly observed standard stars in each filter.
Unfortunately, these stars span a fairly narrow range in color, roughly 0.35 mag $<$ $R-I < 0.50$ mag; the only bluer standard star in the field, SA107-215, has very large photometric errors in Landolt's table, and consequently, it was not used.

Photometry of the GRB 980329 field was done in a similar fashion:  a
curve of growth was derived from multiaperture photometry and used to
correct all magnitudes to an effective 7 arcsec radius, using the IRAF
task ``mkapfile.''  Photometric errors were also taken from
``mkapfile''; these include the uncertainties due to photon
statistics, sky subtraction, and aperture corrections.  The magnitudes
were then corrected for the difference in airmass between the
GRB 980329 and SA107 fields, and for color terms between our filters
and the standard filters used by Landolt.  Since our standard stars
were observed at only a single airmass, we adopted standard extinction
coefficients for Kitt Peak, and we conservatively assumed 
uncertainties in each coefficient equal to that coefficient, 
giving $0.03 \pm 0.03$ mag/airmass for
the $I$ band and $0.08 \pm 0.08$ mag/airmass for the $R$ band.  These
values are IRAF defaults, and agree well with the values of 0.04
mag/airmass ($I$ band) and 0.10 mag/airmass ($R$ band) measured during
1996 November at the WIYN telescope on Kitt Peak (Smith 1997).  We 
measured the color terms to be $(+0.010 \pm 0.07)(R-I - 0.42)$ mag for the
$I$ band and $(-0.049 \pm 0.07)(R-I - 0.42)$ mag for the $R$ band, where 
$R-I = 0.42$ mag is the approximate median color of the observed standard 
stars, and where the signs indicate that blue objects appear brighter in
the $I$ band and fainter in the $R$ band than they would for the standard
filters.  If we fix the color and extinction terms and fit only for
the photometric zero point, we measure an uncertainty of $\pm 0.004$
mag for both filters; this corresponds to the statistical uncertainty
in the photometric calibration arising from uncertainties in the flux
measurements of the our standard stars.

The overall uncertainty in the calibrated magnitude of each measured star
in the GRB 980329 field was computed as the sum in quadrature of that
star's photometric uncertainty, the zero point uncertainty ($0.004$
mag), the systematic uncertainty in the extinction term for
a difference in airmass of $0.3$, and the systematic uncertainty in the 
color term.  For the $I$ band, this is the quadratic sum of 
the photometric error,
$0.004$ mag, $0.010$ mag, and $0.07(R-I - 0.42)$ mag; for the $R$
band, it is the quadratic sum of the photometric error, $0.004$ mag, $0.024$
mag, and $0.07(R-I - 0.42)$ mag.  Since these errors include a substantial 
systematic component, the final magnitude errors for different stars in the
GRB 980329 field are not independent.  We list the calibrated
magnitudes and their final errors for 11 stars in the
GRB 980329 field in Table 2.  A finding chart is shown in Figure 1.

\section{1.34-m Tautenburg Schmidt Telescope Observations}

On March 29, Klose observed the GRB 980329 field with the 1.34-m Tautenburg Schmidt telescope, using a thinned, back-illuminated Tektronix 1024 $\times$ 1024 CCD.  The pixel scale is 1.2 arcsec per pixel, which corresponds to a 20 arcmin field of view.

A series of 27 120-second exposures were taken between March 29.794 - 29.856 UT, nine through the $V$-band filter, nine through the $R$-band filter, and nine through the $I$-band filter.  A second series of 36 120-second exposures were taken between March 29.863 - 29.947 UT, likewise, twelve through each of the $V$-, $R$-, and $I$-band filters.  To avoid problems with hot and warm pixels, the telescope was dithered between exposures.
The sky was nearly photometric at the beginning of the observations; however, observing conditions worsened, and by March 29.95 UT, cloud cover rendered further observations useless.  
During the observations, the airmass ranged from 1.05 to 1.61.

The seven best $I$-band images of the first series and the ten best $I$-band images of the second series were combined using standard IRAF tasks into two images:  the equivalent of an 840 second exposure of mean epoch March 29.827 UT, and the equivalent of a 1200 second exposure of mean epoch March 29.907 UT.  A source is clearly visible at the position of VLA J0702+3850 in both of these images (Klose 1998; Klose, Meusinger, \& Lehmann 1998).  Using the KPNO $I$-band calibration of Table 2, we find that the $I$-band magnitude of the source is $20.8 \pm 0.3$ in both the March 29.827 UT image and the March 29.907 UT image.  The combined March 29.827 $+$ 29.907 UT, $I$-band image is shown in Figure 2.

Observations could not be made on the following night; however, $I$-band observations of the GRB 980329 field were again made between March 31.8 - 31.9 UT.  However, no source was detected at the position of VLA J0702+3850 to a limiting magnitude of $I \approx 21$.  
Likewise, no source was detected at the position of VLA J0702+3850 in the $V$- and $R$-band images taken on March 29 (Klose, Meusinger, \& Lehmann 1998).

\section{APO 3.5-m Telescope Observations}

On April 1, the Astrophysical Research Consortium (ARC) GRB Afterglow Collaboration observed the GRB 980329 field with the ARC's 3.5-m telescope at APO.  
Both optical observations, using the Seaver Prototype Imaging Camera (SPICam), and NIR observations, using the Near Infrared Grism Spectrometer and Imager II (GRIM II), were made. 

Six 10-minute exposures, centered on the position of VLA J0702+3850, were taken with the SPICam between April 1.097 - 1.143 UT through the $R$-band filter.  
SPICam has a thinned SITe 2048 $\times$ 2048 CCD; we used a 2 $\times$ 2 binning, which corresponds to a pixel scale of 0.28 arcsec/pixel.  The effective seeing was 1.0 arcsec.  The effective airmass was 1.042.

Each image was overscan-subtracted and flat-fielded with twilight flats, using standard IRAF tasks.  The images were then combined into a single, stacked image; a high-sigma threshold clipping was applied to reject deviant pixels.  The stacked image is equivalent to a 3600 second exposure of mean epoch April 1.120 UT.  
A source is clearly visible at the position of VLA J0702+3850 (see Figure 3).  
Using the KPNO $R$-band calibration of Table 2, we find that the $R$-band magnitude of the source is $25.35^{+0.35}_{-0.25}$.

The GRIM II has a NICMOS array, with both $J$-band and Mauna Kea $K'$-band (bandpass 1.95 - 2.30 $\mu$m) filters.  At the focal length of $f/5$, the pixel scale is 0.47 arcsec/pixel, which corresponds to a 2 arcmin field of view.  
Between April 1.164 - 1.182 UT, we took randomly-dithered, 15-second frames every 20 seconds of the VLA J0702+3850 field through the $J$-band filter.
We combined these frames into a single, stacked image with a total integration of 945 seconds and a mean epoch of April 1.173 UT.
Between April 1.183 - 1.207 UT, we took randomly-dithered, 20-second frames every 25 seconds through the $K'$-band filter; the stacked image has a total integration of 1310 seconds and a mean epoch of April 1.195 UT.  
We returned to the VLA J0702+3850 field almost two hours later, and again observed through the $J$-band filter:  between April 1.267 - 1.281 UT, we took randomly-dithered, 15-second frames every 20 seconds; the stacked image has a total integration of 915 seconds and a mean epoch of April 1.274 UT.
UKIRT faint standards 14 (B. Zuckerman, private communication) and 33 (Turnshek \etal 1990) were observed throughout the evening through both filters for both magnitude calibration and airmass correction.

Using a 3 arcsec diameter aperture, we find 1 $\sigma$ detection
limits of 20.8 mag for the first $J$-band image, 19.7 mag for
the $K'$-band image, and 20.7 mag for the second $J$-band image.
Combining the first and the second $J$-band images yields a 1 $\sigma$ 
detection limit of 21.3 mag.
We list corresponding 3 $\sigma$ detection limits in Table 1.
No source was detected at the position of VLA J0702+3850 in either the
combined $J$-band image (Cole et al. 1998a,b), or in the $K'$-band image to
these limiting magnitudes.

\section{KPNO Mayall 4-m Telescope Observations}

Once combined, the four 10 minute exposures of the GRB 980329 field with the KPNO Mayall 4-m telescope (see \S2) correspond to the equivalent of a 1200 second exposure of mean epoch April 3.147 through the $R$-band filter, and the equivalent of a 1200 second exposure of mean epoch April 3.167 through the $I$-band filter.  No source is detected at the position of VLA J0702+3850 in either of these images; however, we place upper limits.  The $I$-band upper limit, in particular, is constraining (see Figure 5).

For the $R$-band image, the most constraining upper limit is obtained with a 0.75 arcsec radius aperture; in this case, the aperture-corrected flux (see \S2) of a point source at the position of VLA J0702+3850 is measured to be $0.40 \pm 0.25$ $\mu$Jy.  Photon statistics and sky subtraction both contribute to this uncertainty.
A larger aperture, in this case, of 1.25 arcsec radius, yields a larger aperture-corrected flux:  $0.80 \pm 0.25$ $\mu$Jy.  
This discrepancy could be due to astrometric error, in which case flux would be more accurately measured with the larger aperture; however, this discrepancy is more likely due to some other source of error, perhaps flat fielding error or uncertainty in the curve of growth with which the aperture corrections are made (see \S2).
Accepting as conservative a 3 $\sigma$ upper limit of 1.50 $\mu$Jy gives $R > 23.3$ mag (Rhoads \etal 1998).

For the I band, the aperture-corrected fluxes measured in the 0.75 and 1.25 arcsec radius apertures are $0.06 \pm 0.51$ $\mu$Jy and $-0.12 \pm 0.44$ $\mu$Jy, respectively.  Accepting as conservative a 3 $\sigma$ upper limit of 1.48 $\mu$Jy gives $I > 23.0$ mag (Rhoads \etal 1998).

\section{Keck-I 10-m Telescope Observations}

On April 6 and April 8, Metzger observed the VLA J0702+3850 field with the Keck-I 10-m telescope using the Near-Infrared Camera (NIRC, Matthews \etal 1993) through the $J$- and $K$-band filters.  The NIRC has 256 $\times$ 256 pixels that project to 0.15 arcsec/pixel on the sky, which corresponds to a 38 arcsec field of view.  

On April 6.27 UT and again on April 8.28 UT, 5-second frames were taken through the $K$-band filter in a standard dither pattern.  The frames were flattened and co-added with bad pixel rejection to produce a single image for each night; each image has a total integration of 1080 seconds.
On April 6.30, 10-second frames were taken through the $J$-band filter in a similar fashion.  Again, a single image with a total integration of 1080 seconds was produced.  Several NIR standards were observed on each night, both of which were clear, for magnitude calibration and airmass correction for each band.  Clipped mean stacks of the off-position dither frames were used for the sky estimate in each case.
 
A source is clearly visible at the position of VLA J0702+3850 on each night, and in each band.  The $K$-band magnitude of this source is $21.4 \pm 0.2$ on April 6.27 UT and $21.9 \pm 0.2$ on April 8.28 UT (Metzger 1998a,b).  The $J$-band magnitude of this source is $23.3 \pm 0.4$ on April 6.30 UT.  
Fluxes were measured in a 1.5 arcsec diameter aperture and corrected to an effective 3 arcsec diameter aperture using curves of growth from the brighter stars in the field; airmass corrections were also applied.
We show the combined April 6.27 $+$ 8.28 UT $K$-band image in the right panel of Figure 4; we show the $J$-band image in the left panel of Figure 4.

\section{Some Implications of the Observations}

In Figure 5, we plot the $R$-, $I$-, $J$-, and $K$-band light curves of the source that is coincident with the variable radio source VLA J0702+3850, using the measurements that we have compiled in Table 1.
In this figure, we have corrected these measurements for Galactic extinction, using the dust maps of Schlegel, Finkbeiner, \& Davis (1998).  These maps are reprocessed composites of the {\it COBE}/DIRBE and {\it IRAS}/ISSA maps, with zodiacal foreground and confirmed point sources removed.  With DIRBE-quality calibration and {\it IRAS} resolution, these maps are more than twice as accurate at extinction estimation than the H I maps of Burstein \& Heiles (1982) (Schlegel, Finkbeiner, \& Davis 1998). 
Using the software that is publicly available at their web site\footnote{http://astro.berkeley.edu/davis/dust/index.html}, we find that at the position of VLA J0702+3850, $E(B-V) = 0.073$ mag, which corresponds to $A_V = 0.241$ mag, $A_R = 0.194$ mag, $A_I = 0.141$ mag, $A_J = 0.065$ mag, and $A_K = 0.042$ mag, for $R_V = 3.1$ (Schlegel, Finkbeiner, \& Davis 1998).
For purposes of comparison, we find that the Burstein \& Heiles (1982) maps give $E(B-V) = 0.124$ mag, which corresponds to $A_V = 0.409$ mag, for $R_V = 3.1$.  The 0.168 mag difference in the value of $A_V$ between these two estimates may be due to systematic errors between the dust and H I extinction models of these authors, but it is more likely due to the lower resolution of the Burstein \& Heiles maps:  there appears to be a large dust lane a few arcminutes to the southeast of this position.

Visual inspection of Figure 5 suggests that the source is fading as a power law in these optical and NIR bands.  The light curves do not appear to level off, which suggests that these measurements are not contaminated by an underlying or host galaxy (see \S1); however, such a scenario cannot be ruled out by these data alone.
Assuming a power-law temporal fading, we measure the temporal index, $b$, in each of these bands to be $b_R = -1.28^{+0.18}_{-0.19}$, $b_I \la -1.05$ (3 $\sigma$), $b_J \ga -1.39$ (3 $\sigma$), and $b_K = -0.98^{+0.30}_{-0.28}$.  To within the measured uncertainties, these values are consistent with a single, frequency-independent temporal index of $b = -1.21^{+0.13}_{-0.12}$ ($\chi^2_{min} = 7.064$, $\nu = 10$).  
A similar temporal index is found by in 't Zand \etal (1998b) for the fading X-ray source 1SAX J0702.6+3850, within the error circle of which this optical/NIR source is located.  
This combination of positional coincidence and temporal fading firmly establishes that the optical/NIR source that is coincident with VLA J0702+3850 is the afterglow of GRB 980329.

Finally, we wish to briefly comment upon the form of the optical/NIR spectrum of this GRB afterglow.  As has been noted by Fruchter (1998a) and Reichart \& Lamb (1998), this spectrum is very flat between the $K$ and the $I$ bands, but is dramatically steeper between the $I$ and the $R$ bands.  Fruchter (1998a) proposes that this is the signature of the Lyman-$\alpha$ forest, which would imply that GRB 980329 is at a redshift of $z \sim 5$.  A complete modeling of the radio through X-ray afterglow of GRB 980329 in terms of the relativistic fireball model is presented in Reichart \& Lamb (1998).

\section{Summary of Results}

We report $R$-, $I$-, $J$-, and $K$-band observations of the source that is coincident with the variable radio source VLA J0702+3850, which lies in the error circle of the fading X-ray source 1SAX J0702.6+3850, as well as in the error circle of GRB 980329.
These observations were taken between 15 hours and 10 days after GRB 980329 with the 1.34-m Tautenburg Schmidt telescope, the APO 3.5-m telescope, the KPNO Mayall 4-m telescope, and the Keck-I 10-m telescope.  We find that this optical/NIR source is fading as a power law with a temporal index of $b = -1.21^{+0.13}_{-0.12}$.  This combination of positional coincidence and temporal fading firmly establishes that this source is the afterglow of GRB 980329.  

\acknowledgements

This research was supported in part by NASA grant NAG5-2868 and NASA contract NASW-4690.  M.R.M.'s research was supported in part by Caltech.

\clearpage

\begin{deluxetable}{ccccc}
\footnotesize
\tablecolumns{5}
\tablewidth{0pc}
\tablecaption{Optical and Near Infrared Observations of the Afterglow of GRB 980329}
\tablehead{\colhead{Band} & \colhead{Date\tablenotemark{a}} & \colhead{Magnitude} & \colhead{Telescope} & \colhead{Reference(s)\tablenotemark{b}}}
\startdata
$R$ & Mar 29.99 & $23.6 \pm 0.2$ mag & NTT & 1, 2 \nl
$R$ & Mar 30.93 + 31.87 & $25.0 \pm 0.5$ mag & NOT & 2, 3 \nl
$R$ & Apr 1.12 & $25.35^{+0.35}_{-0.25}$ mag & APO 3.5 m & 4 \nl
$R$ & Apr 2 & $25.7 \pm 0.3$ mag & Keck-II 10 m & 5 \nl
$R$ & Apr 3.15 & $> 23.3$ mag\tablenotemark{c} & KPNO Mayall 4 m & 4, 6 \nl
$I$ & Mar 29.83 + 29.91 & $20.8 \pm 0.3$ mag & Tautenburg Schmidt & 4, 6, 7, 8 \nl
$I$ & Apr 3.17 & $> 23.0$ mag\tablenotemark{c} & KPNO Mayall 4 m & 4, 6 \nl 
$J$ & Mar 29.85 & $> 19.6$ mag\tablenotemark{c} & TIRGO & 2, 9 \nl
$J$ & Apr 1.17 + 1.27 & $> 19.4$ mag\tablenotemark{c} & APO 3.5 m & 4, 10, 11 \nl
$J$ & Apr 6.30 & $23.3 \pm 0.4$ mag & Keck-I 10 m & 4 \nl
$K$ & Apr 1.19 & $> 17.8$ mag\tablenotemark{c} & APO 3.5 m & 4, 10, 11 \nl
$K$ & Apr 2 & $20.7 \pm 0.2$ mag & Keck-I 10 m & 12, 13 \nl
$K$ & Apr 3 & $20.9 \pm 0.2$ mag & Keck-I 10 m & 12, 13 \nl
$K$ & Apr 6.27 & $21.4 \pm 0.2$ mag & Keck-I 10 m & 4, 14, 15 \nl
$K$ & Apr 8.28 & $21.9 \pm 0.4$ mag & Keck-I 10 m & 4, 14, 15 \nl
\enddata
\tablenotetext{a}{1998 March 29.85 - April 8.28, UT in decimal days.}
\tablenotetext{b}{1. Palazzi \etal 1998a; 2. Palazzi \etal 1998b; 3. Pedersen \etal 1998; 4. this paper; 5. Djorgovski \etal 1998; 6. Rhoads \etal 1998; 7. Klose 1998; 8. Klose, Meusinger, \& Lehmann 1998; 9. Mannucci \etal 1998; 10. Cole \etal 1998a; 11. Cole \etal 1998b; 12. Larkin \etal 1998a; 13. Larkin \etal 1998b; 14. Metzger 1998a; 15. Metzger 1998b.}
\tablenotetext{c}{3 $\sigma$, 1-sided confidence interval.}
\end{deluxetable}

\clearpage

\begin{deluxetable}{ccccc}
\footnotesize
\tablecolumns{5}
\tablewidth{0pc}
\tablecaption{KPNO $R$- and $I$-Band Calibrations of the GRB 980329 Field}
\tablehead{\colhead{Object\tablenotemark{a}} & \colhead{Right Ascension\tablenotemark{b}} & \colhead{Declination\tablenotemark{b}} & \colhead{$R$ (mag)\tablenotemark{c}} & \colhead{$I$ (mag)\tablenotemark{c}}}
\startdata
1 & 7:02:39.0 & 38:50:32.7 & $15.7 \pm 1.0$\tablenotemark{d} & $15.3 \pm 0.2$\tablenotemark{d} \nl
2 & 7:02:37.5 & 38:50:33.5 & $15.8 \pm 1.0$\tablenotemark{d} & $15.45 \pm 0.12$\tablenotemark{d} \nl
3 & 7:02:35.1 & 38:50:23.2 & $16.3 \pm 0.3$\tablenotemark{d} & $15.988 \pm 0.036$\tablenotemark{e} \nl
4 & 7:02:40.1 & 38:50:11.8 & $16.966 \pm 0.031$\tablenotemark{e} & $16.647 \pm 0.0155$ \nl
5 & 7:02:39.4 & 38:50:03.1 & $18.443 \pm 0.028$ & $18.093 \pm 0.0175$ \nl
6 & 7:02:38.7 & 38:50:26.9 & $20.655 \pm 0.065$ & $19.473 \pm 0.0623$ \nl
7 & 7:02:36.6 & 38:50:36.3 & $20.386 \pm 0.054$ & $19.331 \pm 0.0478$ \nl
8 & 7:02:36.3 & 38:50:19.8 & $20.514 \pm 0.046$ & $19.625 \pm 0.0459$ \nl
9 & 7:02:38.4 & 38:50:50.7 & $21.288 \pm 0.069$ & $20.026 \pm 0.0711$ \nl
10 & 7:02:51.0 & 38:49:31.0 & $17.659 \pm 0.046$ & $16.668 \pm 0.0399$ \nl
11 & 7:02:50.4 & 38:49:57.1 & $17.966 \pm 0.026$ & $17.539 \pm 0.0161$ \nl
\enddata
\tablenotetext{a}{See Figure 1 for a finding chart.}
\tablenotetext{b}{Epoch J2000.}
\tablenotetext{c}{1 $\sigma$ errors are not independent; extinction and color terms introduce systematic errors (see \S2).}
\tablenotetext{d}{Estimate due to saturation.}
\tablenotetext{e}{Possibly marginally saturated.}
\end{deluxetable}

\clearpage

%Users of this table should be aware that the magnitude uncertainties
%for the different stars include some sources of systematic error
%(airmass and color terms) that are not independent from star to star.
%Also, the first two entries are substantially saturated in R and I;
%the third is substantially saturated in R and perhaps marginally in I;
%and the fourth may be marginally saturated in R.

\clearpage

\figcaption[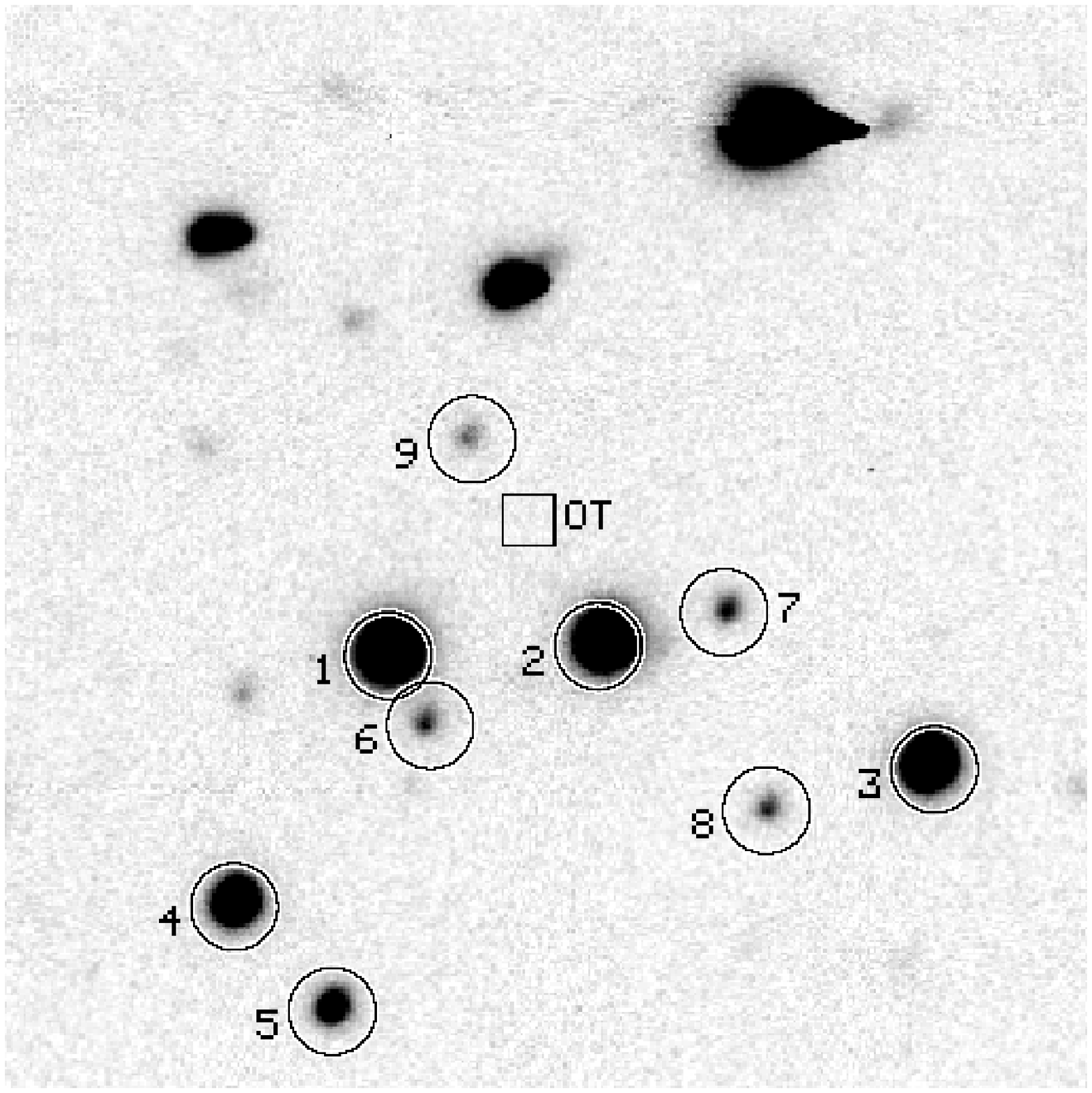]{KPNO Mayall 4-m telescope $I$-band image of the GRB 980329 field (April 3.17 UT).  The image is 90 arcsec square; north is up and east is left.  The optical transient is not detected.  Calibrated stars of Table 2 are labeled.\label{329fig1.eps}}

\figcaption[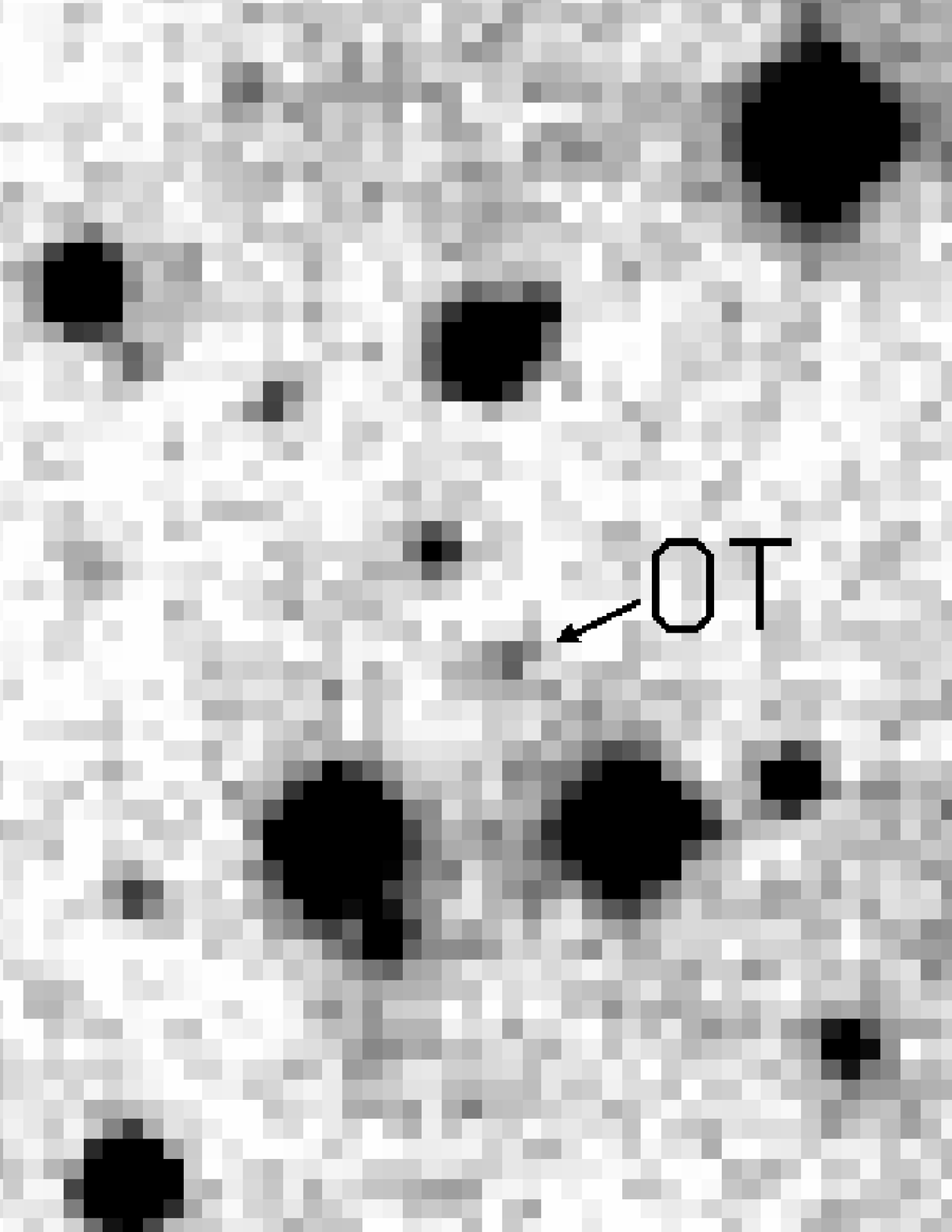]{1.34-m Tautenburg Schmidt telescope stacked (March 29.83 + 29.91 UT) $I$-band image of the GRB 980329 field.  The optical transient is detected (arrow).\label{329fig2.eps}}

\figcaption[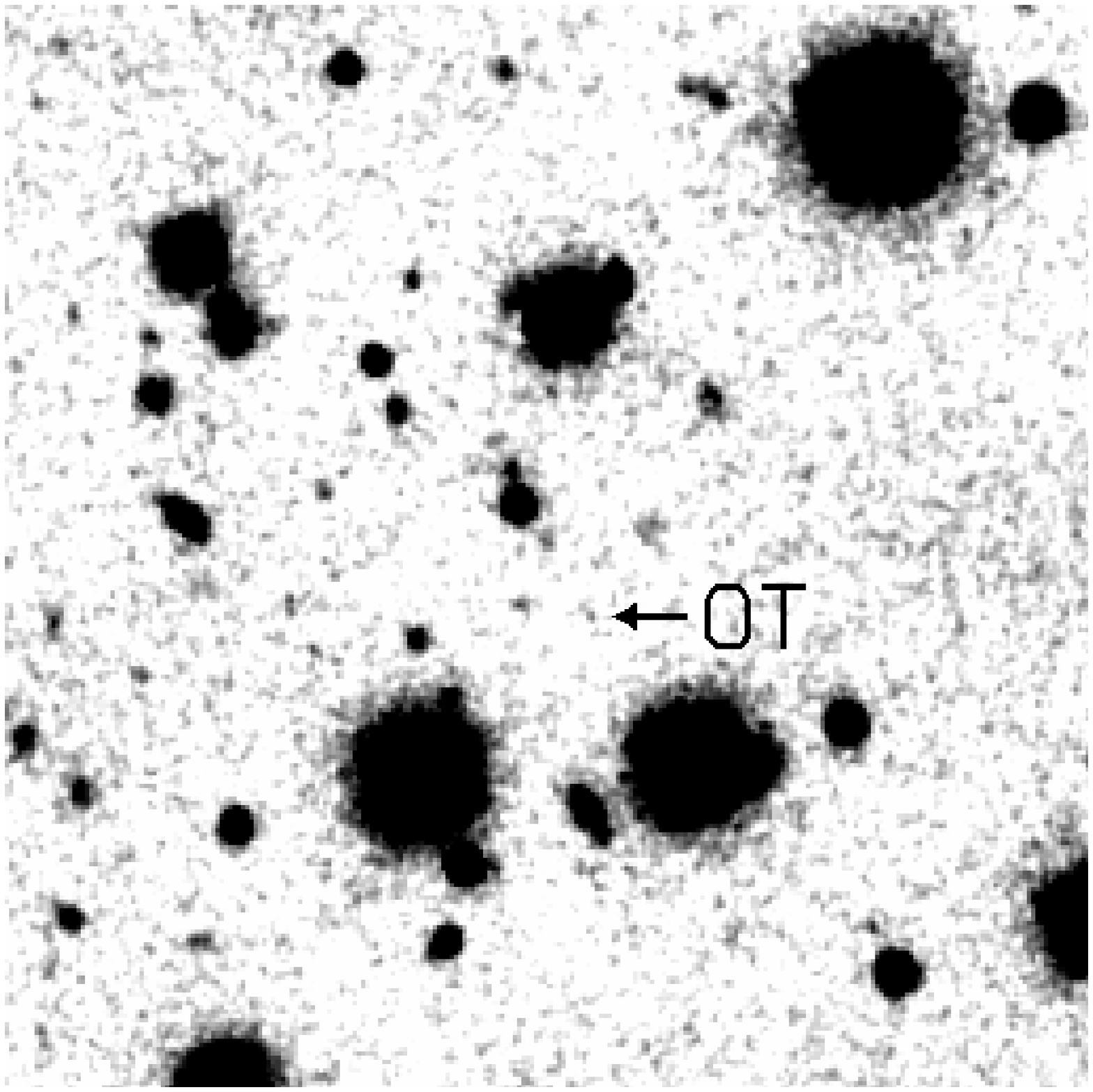]{APO 3.5-m telescope $R$-band image of the GRB 980329 field (April 1.12 UT).  The optical transient is detected (arrow).\label{329fig3.eps}}

\figcaption[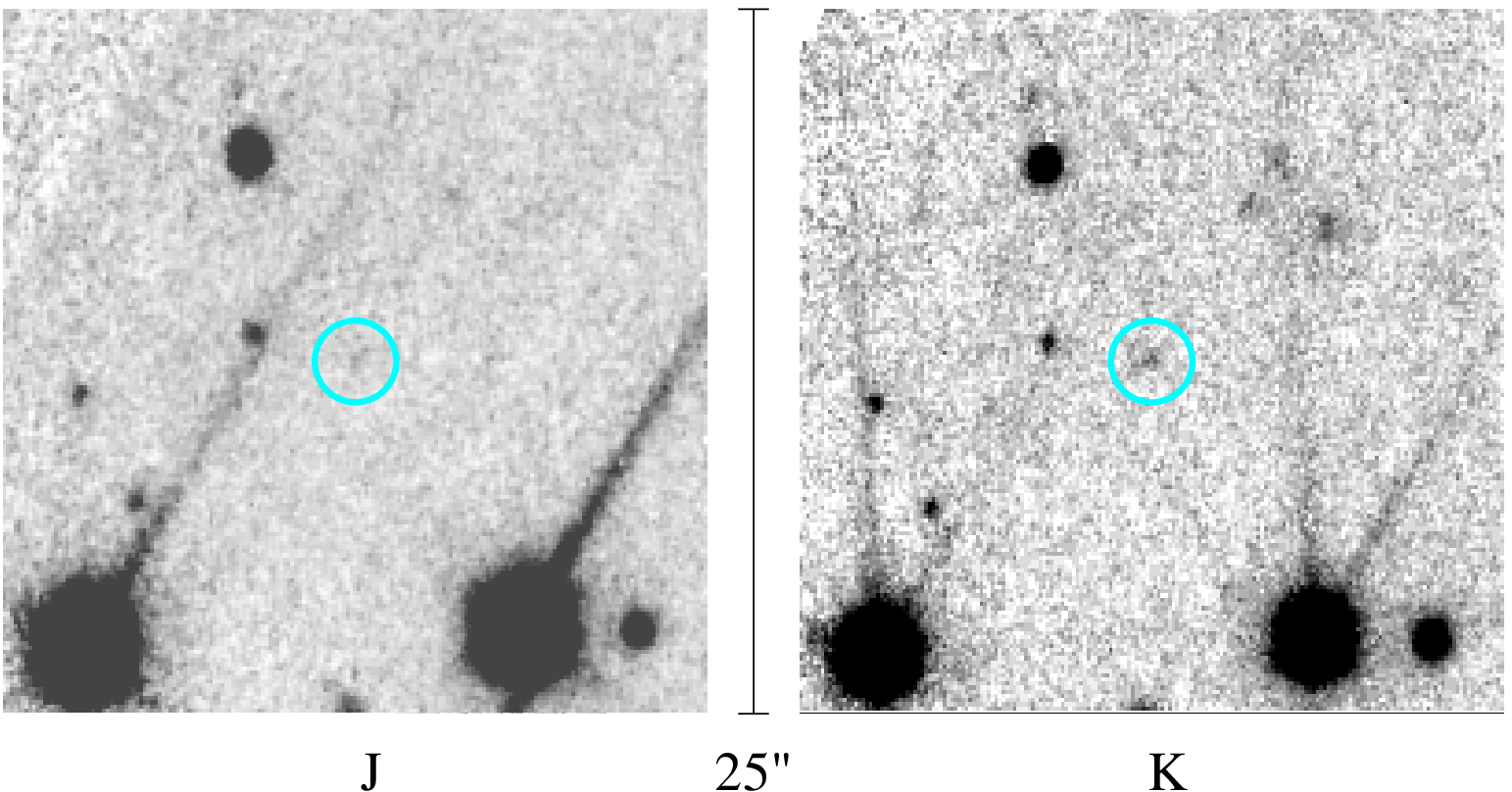]{Keck-1 10-m telescope images of the VLA J0702+3850 field:  $J$-band (April 6.30 UT, left) and stacked (April 6.27 + 8.28 UT) $K$-band (right).  The NIR transient (circled) is detected in both bands, and on both nights (in the $K$-band).\label{329fig4.eps}}

\figcaption[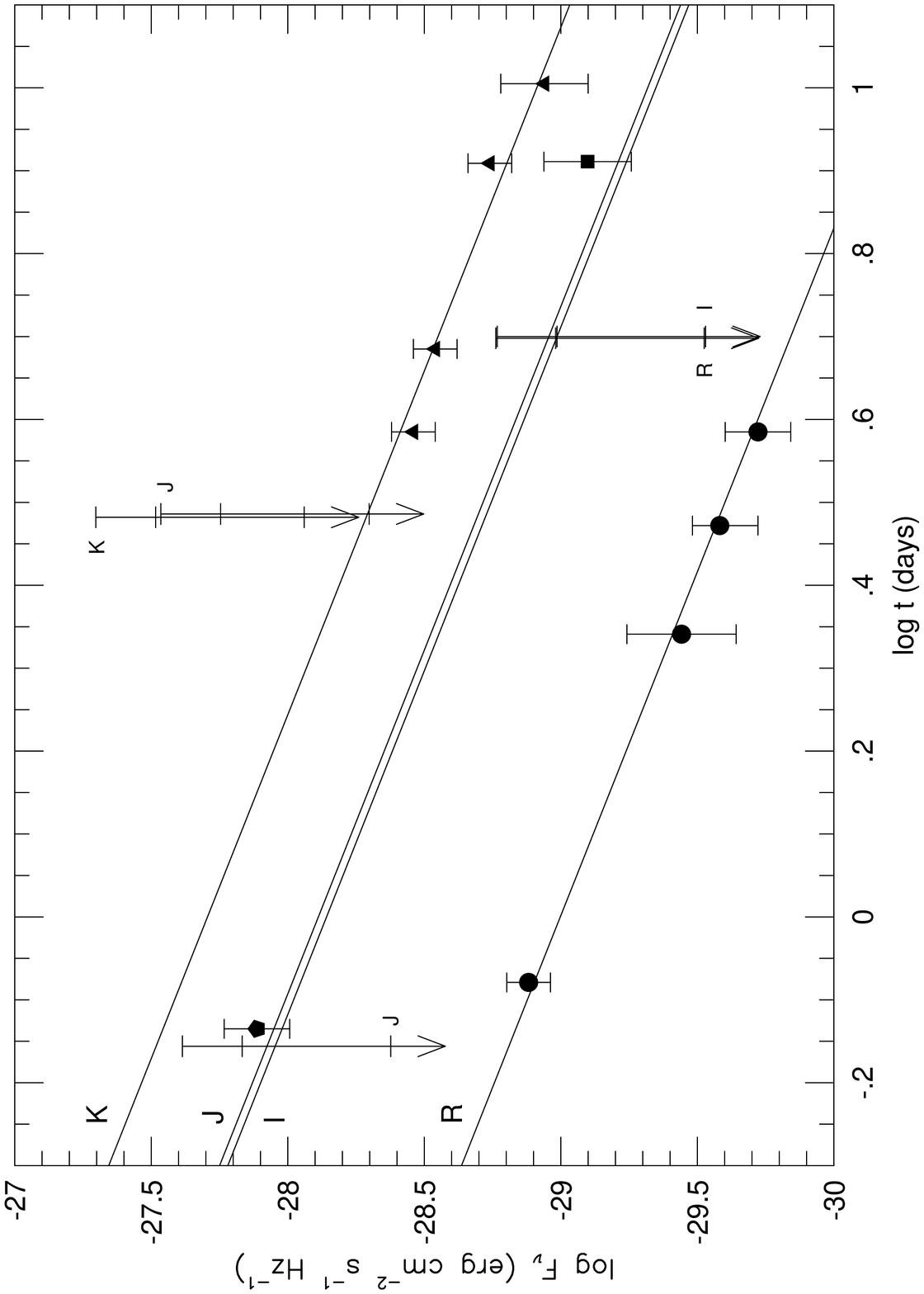]{The $R$-, $I$-, $J$-, and $K$-band light curves of the optical/NIR transient from the measurements of Table 1, corrected for Galactic extinction (see \S7), and the best power law fit to these data.  Circles denote the $R$ band, pentagons denote the $I$ band, squares denote the $J$ band, and triangles denote the $K$ band.  Upper limits are 1, 2, and 3 $\sigma$.\label{329fig5.ps}}

\clearpage

\setcounter{figure}{0}

\begin{figure}[tb]
%\plotone{329fig1.eps}
\plotfiddle{329fig1.eps}{7.5cm}{0}{87.5}{87.5}{-270}{-230}
\end{figure}

\clearpage

\begin{figure}[tb]
%\plotone{329fig2.eps}
\plotfiddle{329fig2.eps}{7.5cm}{0}{55}{55}{-170}{-100}
\end{figure}

\clearpage

\begin{figure}[tb]
\plotone{329fig3.eps}
\end{figure}

\clearpage

\begin{figure}[tb]
%\plotone{329fig4.eps}
\plotfiddle{329fig4.eps}{7.5cm}{90}{130}{130}{+150}{-175}
\end{figure}

\clearpage

\begin{figure}[tb]
\plotone{329fig5.ps}
\end{figure}

\end{document}